\documentstyle[12pt,aaspp4,lscape,flushrt]{article}

\def\kms{km~s$^{-1}$}
\def\cm2{cm$^{-2}$}

\def\lya{{\rm Ly}$\alpha$}

\newcommand{\lsim}{\ \raise -2.truept\hbox{\rlap{\hbox{$\sim$}}\raise5.truept
        \hbox{$<$}\ }}
\newcommand{\gsim}{\ \raise -2.truept\hbox{\rlap{\hbox{$\sim$}}\raise5.truept
        \hbox{$>$}\ }}

\begin{document}

\input epsf

\title{THE METAL ABSORPTION SYSTEMS OF THE HUBBLE DEEP FIELD SOUTH QSO}

\author{Sandra Savaglio\footnote{Present address: 
Space Telescope Science Institute, 3700 San Martin Drive, Baltimore, MD21218, 
USA.}}

\affil{European Southern Observatory, Karl-Schwarzschildstr. 2,
        D--85748 Garching bei M\"unchen, Germany}
\altaffiltext{2}{Based on observations collected at the European
  Southern Observatory, La Silla, Chile (ESO Nr. 60.B-0381).}

\begin{abstract}
The Hubble Deep Field South (HDFS) has been recently selected and the
observations are planned for October 1998.  We present a high
resolution (FWHM $\simeq 14$ \kms) spectrum of the quasar J2233--606
($z_{em}\simeq2.22$) which is located 5.1 arcmin East of the HDFS. The
spectrum obtained with the New Technology Telescope redward of
the Lyman--$\alpha$ emission line covers the spectral range 4386--8270
\AA. This range corresponds to redshift intervals for CIV and MgII intervening
systems of $z=1.83-2.25$ and $z=0.57-1.95$ respectively. The data
reveal the presence of two complex intervening CIV systems at redshift
$z=1.869$ and $z=1.943$ and two complex associated ($z_{abs} \approx
z_{em}$) systems. Other two CIV systems at $z=1.7865$ and $z=2.077$,
suggested by the presence of strong Lyman--$\alpha$ lines in low
resolution ground based and Hubble Space Telescope (HST) STIS
observations (Sealey et al. 1998) have been identified. The system at
$z=1.943$ is also responsible for the Lyman limit absorption seen in
the HST/STIS spectrum.  The main goal of the present work is to provide
astronomers interested in the Hubble Deep Field South program with
information related to absorbing structures at high redshift, which are
distributed along the nearby QSO line of sight. For this purpose, the
reduced spectrum, obtained from three hours of integration time, has
been released to the astronomical community.
\end{abstract}

\keywords{cosmology: observations -- quasars: absorption lines --
quasars: individual: J2233--606}

\section{INTRODUCTION}

The Hubble Deep Field (HDF) program (Williams et al. 1996), which was
carried out with the Hubble Space Telescope (HST) in December 1995 in
the four filters $UBVI$ to the deepest ever reached limiting
magnitude, together with followup observations in other spectral
bands, can be considered as one of the major astronomical events of
the nineties.  More than 40 articles on this program have been
published in refereed journals between April 1996 and December 1997,
confirming the tremendous impact of the observations. Articles related
to the high redshift Universe ($z\gsim1$) treat such subjects as the
galaxy redshift distribution (Gwyn \& Hartwick 1996, Seidel et
al. 1996, Lanzetta et al. 1996, Lowenthal et al. 1997), the Global
Star Formation History (Madau et al. 1996, Connoly et al. 1997, Guzman
et al. 1997) and the clustering properties of galaxies (Colley et
al. 1996, Villumsen et al. 1997).

After two years, the Hubble Deep Field South (HDFS) has been
planned for Cycle 7 of the HST (Williams et al. 1997, see also the
HDF--South Web site at the URL \\
http://www.stsci.edu/ftp/science/hdf/hdfsouth/hdfs.html),
and the observations will be performed in October
1998. Around the selected Southern field only one high redshift quasar
J2233--606 ($z_{em}\simeq2.22$) has been identified (Boyle 1997), 

and it is located 5.1 arcmin EAST of the HDFS. Two HST orbits
have been dedicated to this object to obtain a low resolution STIS
spectrum (3700 seconds with G230L and 2200 seconds with G430L). The
spectrum in the range $\lambda\lambda=1600-5700$ \AA~reveals, among
other features, the presence of a Lyman limit break at $z\sim1.9$, a
few metal lines in the red part of the \lya~emission line and a
\lya~forest with a few absorption lines on the top of the
\lya~emission lines. No Damped \lya~profile is seen along the line of
sight. Ground--based low resolution observations combined with the
STIS spectrum have been analyzed by Sealey et al. (1998). They report
a first tentative identification of metal lines and provide a HI
column density of the Lyman limit system at $z=1.943$, of $N(\rm
HI) = (3.1\pm1.0)\times10^{17}$ \cm2.

Here we report high spectral resolution (FWHM $\simeq 14$ \kms)
observations obtained with the New Technology Telescope (NTT) of the
nearby QSO J2233--606 ($z_{em}\simeq2.22$) in the spectral range
$\lambda\lambda=4386-8270$ \AA.  Even though the 3 hours of
integration time have not produced an extremely high quality spectrum,
the reduced data have been released to the astronomical community (see
the URL http://www.eso.org/ $\tilde{}$ ssavagli). It is hoped that
these data will provide a valuable guide when planning HDFS followup
observations and will assist scientific discussions on the relation
between the absorption systems along the QSO line of sight and the
emitting objects which will be identified in the HDFS.  Other high
resolution observations from the ground primarily covering the
Lyman--$\alpha$ forest region have been scheduled for July 1998 at the
high resolution spectrograph UCLES of the Anglo Australian Telescope
(AAT).

\section{THE DATA}

The spectrum of the quasar J2233--606 ($z_{em}\simeq2.22$) has been
observed in October 1997 at the ESO 3.5m NTT with the high resolution
spectrograph EMMI in the spectral range $\lambda\lambda=4386-8270$
\AA. The log of the three exposures of one hour duration each is reported in
Table \ref{t1}. The flux calibration has been done using the spectrum
of the standard star HR778 (Hamuy et al. 1992) and the flux has turned out to
be about 20\% lower than what was found in the HST/STIS spectrum due to
slit losses. The echelle mode of EMMI can be used only in the red
arm, which does not perform at the highest quality in the blue, where the
interesting part of the spectrum is, between the \lya~and the CIV
emission lines ($\lambda\lambda=3940-5020$ \AA). To avoid overlapping
of the orders in the blue part of the CCD, we limited the slit length
to 4 arcsec, which created some problems for sky subtraction.

Vacuum and heliocentric wavelength calibration obtained using the
Thorium--Argon lamp, gave an accuracy of the order of less than one
tenth of a pixel. The final combined spectrum has been rebinned to a
constant pixel size of 0.1 \AA. The spectral resolution measured from the
lamp emission lines is FWHM $\simeq 11$ \kms, but the real resolution
in the final spectrum is somewhat degraded due to the merging of the
three spectra and to a possible slight drift of the object in the
slit during the exposure. The final assumed resolution  is that
measured from the sky emission lines, and that is FWHM $\simeq 14$
\kms.

Observations were complicated by the low position of the object
($\delta = -60^\circ 33'$) in the sky of la Silla ($\delta = -29^\circ
15'$) and by its visibility.  At the end of October the object was at
the meridian just at the beginning of the night.  The airmass ranged
from 1.2 at the start to 1.5 at the end of the observations. The
seeing monitor station located at La Silla Observatory registered
during the exposures a seeing varying from 0.9 arcsec to 1.5 arcsec,
with an average around 1.1 arcsec. However that should be considered as a
lower limit, since the seeing at the telescope is normally higher
than that measured outside the dome.

Fig.~1 shows the entire spectrum in the spectral range
$\lambda\lambda=4386-8270$ \AA~degraded to low resolution. The
spectrum data reduction -- carried out with the MIDAS software package
ECHELLE -- has been done together with the spectrum of the noise,
which is important for the determination the significance of absorption lines
in the spectrum. That has been calculated from the propagation of the
photon statistics of the object and sky spectra and from the detector
read--out--noise. The redshift of the QSO can be estimated from the
CIV emission line, in spite of the absorption complex close to the
emission peak and the asymmetric profile of the line that complicate
the fit. The peak is probably in the range $\lambda = 4980-5000$
\AA~which gives $z_{em}\simeq2.22$, although as shown by Espey 
(1993) the best estimate of the actual redshift is given by the
low ionization lines, such as the MgII doublet. The obtained redshift
is somewhat lower
than that found by low resolution observations (Sealey et al. 1998),
which give $z_{em} = 2.238$ from the fit of several emission lines.

The MIDAS package FITLYMAN (Fontana \& Ballester 1995) has been used
to measure the parameters of absorption lines. Line fitting through
$\chi^2$ minimization of Voigt profiles has been performed, although
complex features commonly seen in metal systems are difficult to
resolve in this spectrum due to the limited signal--to--noise ratio.
The number of components is normally underestimated due to blending. 
This means that  the fit of a blended profile gives a Doppler
parameter $b$ which is not particularly significant, whereas the column
density corresponds, to a good approximation, to the total column density
of individual real lines.

\section{THE  METAL SYSTEMS OF J2233--606}

To identify metal lines in a quasar spectrum, the first approach is to
look for double features and to compare wavelengths with the most common
and strong doublets typically observed, namely CIV, SiIV and
redshifts MgII. Then isolated lines can be identified if associated with the
discovered systems.  The data presented here cover the CIV, SiIV, MgII
absorptions for intervening systems in the redshift range
$z=1.83-2.25$, $z=2.15-2.25$, $z=0.57-1.95$ respectively.

The spectrum of J2233--606, observed at low resolution with HST/STIS
and from the ground with the Australian National University (ANU) 2.3m
telescope (Sealey et al. 1998), shows a Lyman limit absorption around
2700 \AA~($z\simeq1.94$) with HI column density $N(\rm HI) =
(3.1\pm1.0)\times10^{17}$ \cm2~and a few metal absorption lines
redward of the \lya~emission line identified as possible CIV
absorptions. The high resolution observations presented here, allowed
us to identify the strong CIV doublets, which were seen as single
lines in the STIS and ANU spectra. Given the redshift of these
systems, we have attempted the identification of other absorption
lines in the spectral range associated with every single CIV system
found. We report some detections and in many cases upper limits on ion
column densities.

Tables \ref{t2b}--\ref{t4} list the results of fitting procedures for
the intervening systems. These are three strong CIV systems at
$z=1.7865$, $z=1.869$ and $z=1.943$ and one weak CIV system at
$z=2.077$. One possible MgII system candidate at $z=1.5034$ has been
reported, although it does not show any strong Lyman--$\alpha$ at the
same redshift. We also discuss the system at $z=1.928$ reported by
Sealey et al. (1998), for which we do not find strong absorption
lines.  In Tables \ref{t5} and \ref{t6} we present results for the
associated systems ($z_{abs}\approx z_{em}$) at $z=2.198$ and
$z=2.206$ identified through the detection of CIV absorption.  The
tables are organized as follows: the reported  are relative
to the CIV doublet, or the MgII doublet when CIV is not detected
(i.e. for the system at $z=1.5034$ and for the last two components of
the $z=1.943$ complex system). When an ion has not been detected in
the spectrum, we report only an upper limit to the column density for
a given Doppler parameter $b$ which typically gives a $\chi^2$ for the
residuals of the fit around 2 or larger. In these cases the adopted
Doppler parameter is reported in parenthesis. This is in the range
$20-30$ \kms, considerably higher than that typically observed in
metal absorption lines, which should give conservative column density
upper limits. In a few cases the detected line is given with the Doppler
parameter reported in parenthesis as well, indicating that $b$ has
been assumed rather than obtained (in cases in which the fit has
not given a realistic $b$ value).

The following two sections describe the redshift identification
of the individual intervening and associated metal systems.

\subsection{The intervening systems}

In spite of the fact that the observed spectral range covers a wide
redshift interval for the MgII absorption ($z=0.57-1.95$), one of the
most typical metal doublets observed in QSO spectra, only one strong
MgII doublet has been identified, at $z=1.943$. The observational
limit for the column density of this doublet varies along the spectrum
and for a Doppler parameter of 20 \kms~is $\log N(\rm MgII) \simeq
13.1$ around 4700 \AA, 12.7 around 5500 \AA, 12.5 around 6200
\AA~where the S/N is maximum, 12.6 around 7000 \AA~and 12.9 around
7900 \AA. The SiIV coverage is very limited with no detection. The CIV
range extends from $\sim 4400$ \AA~to the emission line at $\sim 5000$
\AA.

\subsubsection{The metal system at $z=1.5034$}

We report this system through a tentative identification of a weak MgII
doublet at $\sim 7000$ \AA~(Fig.~2) and it needs further observations
to be confirmed. No other lines have been detected (Table~\ref{t2b}).
If this metal system were real, it should have shown a strong
\lya~absorption line at 3043 \AA~which does not appear in the STIS
data. The closest strong feature seen is around 3000 \AA, which would
correspond, in the case of a \lya~line, to a redshift of
$z\simeq1.47$, that is 4000 \kms~blueward of this MgII doublet.

\subsubsection{The metal system at $z=1.7865$}

Sealey et al. (1998) report a possible metal system at redshift
$z=1.787$ associated with a prominent \lya~line seen in the STIS
spectrum and perhaps with a CIV doublet. The associated CIV doublet
would be at $\lambda\simeq4320$ \AA. At that spectral interval, our
EMMI data have very little signal, and for this reason we have
excluded at first that part of the CCD from the data
reduction. Subsequently, we dedicated particular attention to this
interval, in an attempt to detect the CIV absorption. In Fig.~3 we show
the two components found, which give a total CIV column density $\log
N(\rm CIV) \simeq 14.3$ (Table \ref{t2c}).

At $\lambda \simeq7951.5$ \AA~there is an absorption line that
corresponds to the MgI$\lambda\lambda2852$ line of the second
component of this system and that looks like a complex structure. We
have fitted it with one line at $z=1.787115$.  The MgI associated
with the first, stronger CIV component is not detected.

\subsubsection{The metal system at $z=1.869$}

This metal system shows a complex CIV absorption (Fig.~4) and
only a much higher S/N in the spectrum could make it possible to
deblend the lines. We have fitted 5 components, but there could be
more. The total CIV column density is $N(\rm CIV) \simeq
3\times10^{14}$ \cm2~with a large uncertainty due to the uncertainties
on the individual components (Table \ref{t2}). Besides the CIV
complex structure, we identified the MgI$\lambda\lambda2852$ lines at
$\lambda \simeq 8185$ \AA~(Fig.~4). The second CIV component
(at $z=1.867628$) shows a double MgI feature for which we report only
the total column density (in this case $b=5$ \kms~is assumed for both
absorptions).

The presence of low ionization species like MgI is typical for high HI
density clouds, e.g. in damped \lya~systems. For a Lyman limit system
with $\log N(\rm HI) \sim 17$, the MgI column density should be much
lower than our detection limit (for a metallicity of 1/100 of solar)
and at least 2 orders of magnitude lower than the MgII column
density. If the MgI identification were real, we should have also
detected the MgII doublet.  This consideration makes the MgI detection
in this system suspicious.  The other explored possibility is that the
complex absorption at $\lambda\simeq8185$ \AA~corresponds to the
MgII$\lambda\lambda2796$ lines of the system at $z=1.928$, that Sealey
et al. (1998) identified as a possible contributor to the Lyman limit
break. In the following paragraph we argue that this solution is also not
satisfactory, making the identification of the absorption lines
seen at $\lambda\simeq8185$ \AA~still doubtful.

The spectrum shows at $\lambda=6722.2$ \AA~an absorption line which
would correspond to the FeII$\lambda\lambda$2344 feature at
$z=1.8676$, but the fit -- with all the 6 lines of the FeII ion in the
spectral range -- rules out this identification.

\subsubsection{A possible metal system at $z=1.928$}

Although Sealey et al. investigate this system as a candidate
contributor to the observed Lyman break at $\lambda\simeq2700$ \AA,
they in fact conclude that this is probably dominated by the system at
$z=1.943$. Table \ref{t3a} gives the  column density due to the
identification of the $\lambda\lambda 2484$ line, and the FeII
column density due to the identification of the
FeII$\lambda\lambda2382$ line. Both of these lines are very weak and
their detection needs further confirmation. No associated CIV has been
found and the upper limit to the column density is $\log N(\rm CIV)
<13.7$. Another possible detection would be the
MgII$\lambda\lambda2796$ at $\lambda \simeq 8185$ \AA, but we consider
the identification unlikely since the second component of the doublet
MgII$\lambda\lambda2803$ is not detected.

\subsubsection{The metal system at $z=1.943$}

This system is the main contributor to the Lyman break seen in the
STIS spectrum at 2700 \AA~($\log N(\rm HI) \simeq17.5$). It is
probably associated with multi--phase gas clouds since it shows a
strong CIV doublet with a double structure and a MgII doublet with a
triple structure  by $100-200$ \kms~(Fig.~5). The S/N is not
high enough to resolve a higher number of components which are often
seen in strong CIV systems. The results of the fitting procedures are
shown in Table \ref{t3}. The first CIV component at $z=1.942007$ is
not saturated and it gives a reasonable result for $\log N(\rm CIV)
\simeq 14.1$. The second at $z=1.942582$ is saturated with a higher
uncertainty on the fitted parameters. The total CIV column density is
$\log N(\rm CIV) = 14.7\pm0.5$. The MgII column density might indicate
a high metallicity compared with that found in typical Lyman limit
systems (in the range between 1/100 and 1/30 of solar, Steidel
1990). That would be confirmed by the detected AlIII ion in the second
component of the system, but not by the low upper limit found for
the column density of the AlII ion.

There is a double feature corresponding to CI at $z=1.943607$ and
$z=1.944033$ at $\lambda\simeq4594$ \AA~which could be associated with
the two reddest MgII components. Unfortunately the signal in the
spectrum is very low and the identification is doubtful.

\subsubsection{The metal system at $z=2.077$}

This metal system shows a very weak CIV doublet and that is the only
detected metal ion, whereas the STIS spectrum shows a strong \lya~line
at the same redshift. The CIV$\lambda\lambda1550$ line falls very
close to a noisy peak making the detection of this system more
complicated. The CIV fit requires at least two components. The results are
given in Table \ref{t4} and shown in Fig.~6 assuming $b=5$ \kms~for
both the components. The observed total CIV column density of a few
$10^{13}$ \cm2~could give an HI column density absorption around
$10^{15}$ \cm2~for a cloud with a metallicity of 1/100 of solar,
photoionized by a UV background QSO dominated with a softness
parameter $S_L \equiv J_{912}/J_{228} \sim 100$\footnote{The softness
parameter $S_L$ is defined as the ratio of the intensities of the UV
background $J_\nu$ at the HI and HeII Lyman limit, at 912 \AA~and 228
\AA~respectively.} and absorbed by the Intergalactic Medium.

\subsection{The associated systems}

The STIS spectrum of J2233--606 shows several absorption lines around
the peak of the \lya~emission line, some of which are \lya~lines
with CIV absorption. Fig.~7 shows a portion of the EMMI
spectrum smoothed over 9 pixels around the CIV emission peak (at
$z\simeq2.22$) in the redshift range $\pm5000$ \kms. A cluster of CIV
lines is clearly seen around 4960 \AA. Other spurious features are
visible, but no other CIV doublets have been identified, although they
are probably present.

Associated absorption systems ($z_{abs}\approx z_{em}$) have caught
the astronomers' interest since they give the opportunity to explore the
physical conditions of high redshift absorbing gas clouds. It is not
clear yet whether these clouds are associated with the quasar itself,
have their origin in the Interstellar Medium of the hosting objects or
are located somewhere in the cluster which contains the quasar. It has
been shown (Savaglio et al. 1994, M{\o}ller et al. 1994, Petitjean et
al. 1994, Savaglio et al. 1997, Hamann et al. 1997) that the gas in
these clouds is metal rich already by redshift $z\sim4$, indicating a
strong stellar activity in the environment, also confirmed by an
overabundance of $\alpha$--elements. To compare the properties of the
prominent associated absorption systems in J2233--606 with what is known
so far, we collected high resolution data from the literature (Table
\ref{t2a}) and plotted them in Fig.~8.

Tables \ref{t5} and \ref{t6} list column densities for the systems
found in J2233--606. They basically contain CIV detections and upper
limits for other ions. For completeness we also report results for low
ionization species which fall in the observed range even if they are
typically not detected in the associated systems.

\subsubsection{The metal system at $z=2.1982$}

This associated system (Table \ref{t5}) shows a strong CIV doublet at
about $-2000$ \kms~from the emission peak. The fit with one component
gives a relatively poor result since the equivalent width ratio of the
first line CIV$\lambda\lambda1548$ relative to the second
CIV$\lambda\lambda1550$ is lower than expected (Fig.~9). A fit with a
double structure does not result in a better fit. There may be two
possible explanations for this discrepancy: either the second line is
contaminated by another unidentified absorption line or the level of
the continuum has been underestimated for the first line and/or
overestimated for the second. However, we consider it unlikely that
the second possibility would affect the fit so much.

Other strong doublets typically seen in associated systems are NV and
SiIV. The ratio of the SiIV to CIV column densities is typically spread
over a wide range (Fig.~8) and the upper bound of the distribution is
$\log [N(\rm SiIV)/N(\rm CIV)] \approx -0.5$, which would give for this
system -- with $\log N(\rm CIV) \simeq 13.8$ -- an upper limit to the SiIV
column density of $\log N(\rm SiIV) \lsim 13.3$, just at the level of our
detection limit. Other high or intermediate ionization species are out
of our spectral range. From the correlation seen in the $\log [N(\rm
NV)/N(\rm CIV)]~vs.~\log N(\rm CIV)$ plot (Fig.~8), we expect for the
NV column density at $\lambda \simeq 3966$ \AA~a value which is $\log
N(\rm NV) \approx 13.4$. Concerning the HI column density, the bulk of
the associated systems around the observed CIV column density of $\log
N(\rm CIV) \simeq 13.8$ is $-1.0 < \log [N(\rm CIV)/N(\rm HI)] <$ 0.1
(Fig.~8), which means that we could expect a HI column density in the
range 13.7 $\lsim \log N(\rm HI) \lsim$ 14.8.

\subsubsection{The metal system at $z=2.2058$}

This system shows a multiple component CIV absorption in the redshift
range between about $-1400$ \kms~and $-1100$ \kms~from the emission
peak (Fig.~10). The fitting procedures are very complicated due to a
probable intrinsic blending of the lines and to the low S/N. The
number of components is then very uncertain. In Table \ref{t6} we
report the results for a structure with 6 components for a total CIV
column density around $10^{15}$ \cm2. Upper limits to the SiIV column
density are not particularly significant.  Given the CIV column
density of individual components, the associated HI column densities are
most likely lower than $10^{15}$ \cm2, while NV could be detected with
column density $\log N(\rm NV)\approx 13$.

\section{DISCUSSION}

Unlike in the case of the HDF North program, ``follow--up''
observations of the HDFS region from the ground and space telescopes
began already long before the beginning of the program. The
observations from the radio to X--rays, from low and high resolution
spectroscopy to narrow and broad band imaging, programmed since
November 1997 to the Spring of 1999, will be combined to provide
astronomers with the deepest and most complete view of the Universe
for redshifts $z>1$, where most of the stars have probably already formed.

Moreover, unlike the Northern field, the HDFS is close to a high
redshift quasar.  The QSO line of sight ($\alpha = 22^h 33' 37.67''$,
$\delta = -60^\circ 33' 28.95"$, J2000 Equinox) is located $5'7''$
away from the Hubble Deep Field South ($\alpha = 22^h 32' 56.22''$,
$\delta = -60^\circ 33' 02.69"$, J2000 Equinox). This corresponds to a
real separation that is shown in Fig.~11 in the redshift range $z=0.5-3$. The
separation at the QSO redshift ($z_{em}\simeq2.22$) is 2.8 and 8.9
$h_{50}^{-1}$ Mpc ($h_{50}$ is the Hubble constant
expressed in units of 50 \kms~Mpc$^{-1}$) for a flat Universe in the
physical and comoving space respectively.

Deep imaging of the region near the QSO line of sight is fundamental
to the understanding the quasar environment, the origin of high
redshift quasars in connection to that of galaxy clusters and the
relation between \lya~absorbers and high redshift galaxies. The
typical wavelength of density fluctuations whose collapse generate
galaxy clusters is around 20 $h_{50}^{-1}$ Mpc. On the other hand,
once virialized, clusters have a typical size (i.e., virial radius) of
about $2-5~h_{50}^{-1}$ Mpc, depending on their mass and on the
background cosmological model (e.g., Kitayama \& Suto 1996, and
references therein).  At redshift $z\sim 2.2$ we generally expect to
find protoclusters which are not yet virialized. Instead, they should
correspond to detectable overdensities involving scales of about 10
$h_{50}^{-1}$ Mpc comoving. Therefore, that would roughly correspond
to the comoving separation between the HDFS and the main absorption
clusters seen in the QSO spectrum, whereas at the same redshift, the
comoving HDFS size is around 4.5 $h_{50}^{-1}$ Mpc. The size of the
HFDS and the separation from the quasar might be suitable for the
identification of clustered structures associated with the QSO
absorption systems.

The idea of the existence of clustered structures at high redshifts has
recently been confirmed by observations. Steidel et al. (1998)
found evidence for a protocluster of galaxies at $z\simeq 3.1$ in the
field of two high redshift QSOs, one of which is at the same redshift
as the structure. The discovered 15 Lyman break galaxies are
distributed on a plane which is at least $20\times15~h_{50}^{-2}$
Mpc$^2$ comoving. Numerical simulations have shown that non--linear
structures can be present on large scales already by $z=3$. Using
N--body simulations and semi--analytical methods, Governato et
al.~(1998) were able to produce in the generic Cold Dark Matter
scenario, structures similar to the observed one.  These high redshift
associations of Lyman break galaxies are strongly biased with respect
to the already overdense local dark matter distribution and are the
seeds of present day galaxy clusters. 

Other protoclusters of galaxies candidates at very high redshift
($z\sim2$) have been found in a limited number of cases (Dressler et
al. 1993, Francis et al. 1996, Pascarelle 1996, Hutchings 1995). At
low redshift ($z<1$) extensive imaging and spectroscopic observations
of galaxy clusters in quasar environments and other AGNs indicated
that quasar activity is strongly correlated with the properties of the
environment of the host galaxy (Yee \& Green, 1987; Yee\& Ellingson,
1993). The study of the HDFS QSO should allow to extend this concept
to much higher redshifts. In particular, it should reveal if the
highest density peaks that give origin to quasars at high redshift are
also the seeds of structure formation on much larger scales.

\section{SUMMARY}

We provide the first results of high spectral resolution observations
of the quasar J2233--606 which lies close to the selected Hubble Deep
Field South. The main motivation for this work is to give the
astronomical community interested in the HDFS access to information on
collapsed structures along the QSO line of sight. The spectrum shows
two strong CIV systems, at $z=1.869$ and $z=1.942$, the latter being
also responsible for the Lyman limit absorption seen in the HST/STIS
spectrum. Also detected are two strong associated systems
($z_{abs}\approx z_{em} \approx 2.2$) with prominent CIV
absorption. Moreover, we have confirmed the presence of two other less
certain CIV systems at $z=1.7865$ and $z=2.077$, suggested by the
analysis of the HST/STIS spectrum. Finally one tentative MgII
detection at $z=1.5034$ needs further observations for
confirmation. Despite the limited integration time, we have been able
to perform fitting procedures to the data and derive column density
measurements and upper limits for all the ions in the observed
range. This information can be combined with high resolution
observations of the \lya~forest planned for July 1998 at the
AAT/UCLES, in order to derive the ionization state and the heavy
element enrichment of the identified systems. The ultimate goal of
this study would be to understand the origin and the evolution of the
high redshift structures which will be detected in the Hubble Deep
Field South.

\acknowledgments 
It is my pleasure to thank Stefano Borgani, Adriano
Fontana, Mario Livio, Nino Panagia and John Webb for useful
discussions. I am also grateful to Sandro D'Odorico and Emanuele
Giallongo for support with the observations. My special acknowledgment
goes to the HDFS team, in particular to Massimo Stiavelli for
inspiring this work and to Marcella Carollo and Harry Ferguson for
reducing the STIS data and making them available.

\newpage

\begin{table}
\caption[t1]{Log of the observations.}
{\label{t1}}
\begin{center}
\begin{tabular}{cccccccc}
\hline\hline&&&&&&&\\[-5pt] 
\#    & date  & range  & grating & CD    & slit & exposure & mean \\
      &       & (\AA) &         & grism & width& (s) & airmass\\
[2pt]\hline&&&&&&&\\[-8pt] 
1  & 29/10/97 & $4386-8270$  & 10 & \#3 & 1.2'' & 3600 & 1.2 \\ 
2  & 29/10/97 & $4386-8270$  & 10 & \#3 & 1.2'' & 3600 & 1.3 \\
3  & 29/10/97 & $4386-8270$  & 10 & \#3 & 1.4'' & 3600 & 1.4 \\
[2pt]\hline\\
\end{tabular}\end{center}
\end{table}

\begin{table}
\caption[t2]{The metal system at $z=1.5034$.}
{\label{t2b}}
\begin{center}
\begin{tabular}{l|cc}
\hline\hline&&\\[-8pt]
\multicolumn{1}{c|}{Ion} 
& \multicolumn{2}{|c}{$z=1.503369$} \\
[2pt] & $\log N$ & $b$   \\
\hline\hline&&\\[-8pt]
SiI   & $<13.1$  & (25)   \\
AlIII & $<12.9$  & (25) \\
MgI   & $<11.8$  & (25)   \\
MgII  & 12.37$\pm$0.10 & 22.7$\pm$6.1 \\
FeI   & $<12.4$  & (20) \\
FeII  & $<12.9$  & (20) \\
[2pt]\hline
\end{tabular}\end{center}
\end{table}

\begin{table}
\caption[t5]{The metal system at $z=1.7865$.}
{\label{t2c}}
\begin{center}
\begin{tabular}{l|cc|cc}
\hline\hline&&&&\\[-8pt]
\multicolumn{1}{c|}{Ion} 
& \multicolumn{2}{|c}{$z=1.786142$} 
& \multicolumn{2}{|c}{$z=1.787161$} \\
[2pt] & $\log N$ & $b$ & $\log N$ & $b$  \\
\hline\hline&&\\[-8pt]
CIV   & 14.18$\pm$0.11 & 30.1$\pm$6.3 & 13.82$\pm$1.10 & 5.1$\pm$5.5 \\
SiI   & $<13.3$  & (20)& $<13.3$  & (20) \\
SiII  & $<14.6$  & (20)& $<14.6$  & (20)  \\
AlII  & $<12.5$  & (20)& $<12.5$  & (20) \\
AlIII & $<12.8$  & (20)& $<12.8$  & (20)  \\
MgI   & $<12.1$  & (20)& 12.31$\pm$0.07  & 26.5$\pm$4.3 \\
MgII  & $<12.7$  & (20)& $<12.7$  & (20) \\
FeI   & $<12.8$  & (20)& $<12.8$  & (20) \\
FeII  & $<12.7$  & (20)& $<12.7$  & (20) \\
ZnII  & $<12.7$  & (20)& $<12.7$  & (20)  \\
[2pt]\hline
\end{tabular}\end{center}
\end{table}

\begin{table}
\caption[t3]{The metal system at $z=1.869$.}
{\label{t2}}
\epsfysize=30cm
\centerline{\epsffile{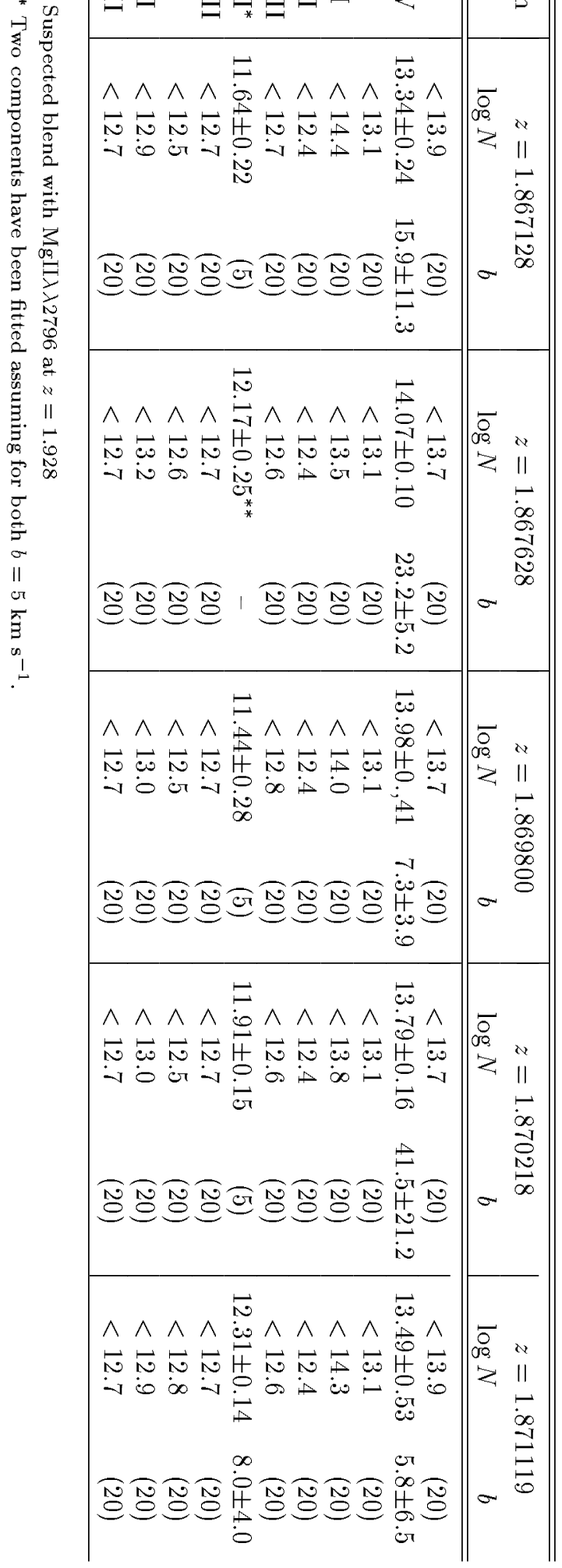}}
\end{table}

\begin{table}
\caption[t5]{The metal system at $z=1.928$.}
{\label{t3a}}
\begin{center}
\begin{tabular}{l|cc}
\hline\hline&&\\[-8pt]
\multicolumn{1}{c|}{Ion} 
& \multicolumn{2}{|c}{$z=1.9288$} \\
[2pt] & $\log N$ & $b$  \\
\hline\hline&&\\[-8pt]
CI    & $<13.8$  & (20)   \\
CIV   & $<13.7$  & (20)   \\
SiI   & $<13.2$  & (20)   \\
SiII  & $<13.6$  & (20)   \\
AlII  & $<12.5$  & (20)   \\
AlIII & $<12.8$  & (20)   \\
MgI   & $<13.3$  & (20)   \\
MgII  & $<12.8$ & (20) \\
FeI   & 12.39$\pm$0.10 & 10.3$\pm$4.5   \\
FeII  & 12.40$\pm$0.16 & (5) \\
ZnII  & $<12.7$  & (20)   \\
[2pt]\hline
\end{tabular}\end{center}
\end{table}

\scriptsize
\begin{table}
\caption[t4]{The metal system at $z=1.943$.}
{\label{t3}}
\begin{center}
\begin{tabular}{l|cc|cc|cc|cc}
\hline\hline&&&&&&&&\\[-8pt]
\multicolumn{1}{c|}{Ion} 
& \multicolumn{2}{|c}{$z=1.942007$}
& \multicolumn{2}{|c}{$z=1.942582$} 
& \multicolumn{2}{|c}{$z=1.943143$}
& \multicolumn{2}{|c}{$z=1.944293$} \\
[2pt] & $\log N$ & $b$  & $\log N$ & $b$ & $\log N$ & $b$& $\log N$ & $b$\\
\hline\hline&&&&&&&&\\[-8pt]
CI    & $<13.7$  & (20) & $<13.7$  & (20) & 13.35$\pm$0.27 &
8.9$\pm$10.3 & 13.34$\pm$0.29 & 14.4$\pm$11.2 \\
CIV   & 14.06$\pm$0.11 & 17.1$\pm$2.7 & 14.61$\pm$0.52 & 16.2$\pm$5.1
& $<13.4$ & (20) & $<13.5$ & (20) \\
SiI   & $<13.3$  & (20) & $<13.3$  & (20) & $<13.3$  & (20)  
& $<13.3$  & (20)   \\
SiII  & $<13.4$  & (20) & $<13.4$  & (20) & $<13.4$  & (20) 
& $<13.4$  & (20)  \\
AlII  & $<12.2$  & (20) & $<12.2$  & (20) & $<12.2$  & (20) 
& $<12.2$  & (20)  \\
AlIII & $<12.6$  & (20) & 12.10$\pm$0.19 & 5.8$\pm$5.8 & $<12.6$  &
(20) & $<12.6$  & (20) \\
MgI   & $<13.1$  & (20) & $<13.1$  & (20) & $<13.1$  & (20) & $<13.1$
& (20)  \\
MgII  & $<12.7$  & (20) & 13.37$\pm$0.14 & 8.8$\pm$1.6 &
12.87$\pm$0.09 & 12.4$\pm$4.1 & 12.74$\pm$0.10 & 8.4$\pm$3.0  \\
FeI   & $<12.6$  & (20) & $<12.6$  & (20) & $<12.7$  & (20) 
& $<12.7$  & (20)  \\ 
FeII  & $<13.0$  & (20) & $<13.0$  & (20) & $<13.0$  & (20)  & $<13.0$
& (20)   \\
ZnII  & $<12.6$  & (20) & $<12.6$  & (20) & $<12.6$  & (20) & $<12.6$
& (20)  \\
[2pt]\hline
\end{tabular}\end{center}
\end{table}
\normalsize

\begin{table}
\caption[t5]{The metal system at $z=2.077$.}
{\label{t4}}
\begin{center}
\begin{tabular}{l|cc|cc}
\hline\hline&&&&\\[-8pt]
\multicolumn{1}{c|}{Ion} 
& \multicolumn{2}{|c}{$z=2.077301$}
& \multicolumn{2}{|c}{$z=2.077561$} \\
[2pt] & $\log N$ & $b$  & $\log N$ & $b$ \\
\hline\hline&&&&\\[-8pt]
CI    & $<13.8$  & (20) & $<13.8$  & (20)  \\
CIV   & 12.95$\pm$0.25 & (5) & 13.08$\pm$0.20 & (5)\\
SiI   & $<12.9$  & (20) & $<12.9$  & (20)  \\
SiII  & $<13.3$  & (20) & $<13.3$  & (20)  \\
AlII  & $<12.3$  & (20) & $<12.3$  & (20)  \\
AlIII & $<12.6$  & (20) & $<12.6$  & (20) \\
MgI   & $<13.4$  & (20) & $<13.4$  & (20)  \\
FeI   & $<13.0$  & (20) & $<13.0$  & (20)  \\ 
FeII  & $<13.0$  & (20) & $<13.0$  & (20)  \\
ZnII  & $<12.7$  & (20) & $<12.7$  & (20)  \\
[2pt]\hline
\end{tabular}\end{center}
\end{table}

\scriptsize
\begin{table}
\caption[t6]{The associated ($z_{abs} \approx z_{em}$) QSO metal systems
known from the literature.}
{\label{t2a}}
\begin{center}
\begin{tabular}{ccccccc}
\hline\hline&&&&&&\\[-8pt] 
QSO    & $z_{abs}$  & $\log N(\rm HI)$ & $\log N(\rm SiIV)$ & $\log
N(\rm CIV)$ & $\log N(\rm NV)$ & ref \\
[2pt]\hline&&&&&&\\[-8pt] 
0000--263&  4.06064 &14.27 &     12.49   &  13.24 & $<12.9$ &Savaglio et al. 1997\\  
	 &  4.06157 &14.48 &     $<12.5$ &  13.47 & $<12.9$ &\\  
	 &  4.06248 &13.19 &     $<12.6$ &  13.26 & $<12.9$ &\\  
	 &  4.10106 &14.40 &     12.39   &  13.21 & $<12.7$ &\\ 
	 &  4.12605 &13.76 &     $<12.7$ &  13.54 & $<12.8$ &\\   
	 &  4.12688 &12.85 &     $<12.5$ &  13.17 & $<12.7$ &\\   
	 &  4.12983 &15.02 &     $<12.5$ &  13.46 & 12.66 &\\   
	 &  4.13111 &15.80 &     12.75   &  14.34 & 13.15 &\\ 
 	 &  4.13245 &15.00 &     12.49   &  14.62 & 13.15 & \\  
	 &  4.13331 &14.56 &     12.40   &  13.68 & $<12.8$ &\\
	 &  4.13420 &15.39 &     12.98   &  14.51 & 13.53 &\\  
[2pt]\hline&&&&&&\\[-8pt]
0123+257& 2.3586 &14.7 &13.6 &   14.7 & 14.8 & Barlow\&Sargent 1997 \\
	& 2.3618 & 14  &--    &   13.7& 13.65 & \\
[2pt]\hline&&&&&&\\[-8pt]
0424--131& 2.100  &   15.3&  12.92 &   13.67 & $<12.7$ & 
Petitjean et al. 1994\\
	& 2.133  &  14.5 &  --     &   13.64& 13.99 & \\ 
	& 2.173  &  14.9 & 12.67  &  14.2 & -- & \\
0450--131& 2.2310 &  14.7 & 13.3   &  13.7 & 14.1 &\\ 
[2pt]\hline&&&&&&\\[-8pt] 
1422+231& 3.5353  &16.36 &13.61  & 14.28& $<12.7$ & Songaila \& Cowie 1996 \\
	& 3.5862  &15.70 &12.32  & 13.29& $<12.3$ & \\
[2pt]\hline&&&&&&\\[-8pt] 
1946+769 &  3.0483 & 13.98 & 13.00 & 13.75 & -- & Fan \& Tytler 1994 \\
	 &  3.0495 & 17.64 & 13.97 &   14.55& 13.32 \\
	& 3.0504 &  14.09 & 13.6  &   14.35 & 12.90 \\
[2pt]\hline&&&&&&\\[-8pt]
2116--358& 2.30709 & 13.7 &  --   &     13.9 & 13.48 & Wampler et al. 1993\\ 
	& 2.30680 & 13.3 & -- & 14.3& 14.6 & \\ 
	& 2.30657 & 13.3 & -- & 13.18& 13.48 &\\ 
	& 2.30625 & 13.58 & -- & 13.85& 13.3 &\\ 
	& 2.30606 & 13 & -- & 12.70& 13.3 & \\ 
	& 2.31880 & 14.18 &  13.78 & 15& 12.9 & \\ 
	& 2.31843 & 17.60 & 13.78 & 14.7& 13.6 \\ 
	& 2.31813 & 14.70 & 13.3 & 14& 13 & \\ 
	& 2.31769 & 14.18 & 12.7 & 14.08& 13.48 & \\ 
	& 2.31702 & 13.9 & 13 & 13.65& 13.48 &\\ 
[2pt]\hline\\
\end{tabular}\end{center}
\end{table}
\normalsize

\begin{table}
\caption[t7]{The metal system at $z=2.1982$.}
{\label{t5}}
\begin{center}
\begin{tabular}{l|cc}
\hline\hline&&\\[-8pt]
\multicolumn{1}{c|}{Ion} 
& \multicolumn{2}{|c}{$z=2.198223$} \\
[2pt] & $\log N$ & $b$   \\
\hline\hline&&\\[-8pt]
CI    & $<13.4$  & (25) \\
CIV   & 13.77$\pm$0.05 & 19.8$\pm$2.4 \\
SiI   & $<13.1$  & (25)   \\
SiII  & $<13.3$  & (25)   \\
SiIV  & $<13.3$  & (25)  \\
AlII  & $<12.3$  & (25)  \\
AlIII & $<12.6$  & (25) \\
MgI   & $<13.3$  & (25) \\
FeI   & $<12.6$  & (20) \\
FeII  & $<13.4$  & (20) \\
ZnII  & $<12.7$  & (20) \\
[2pt]\hline
\end{tabular}\end{center}
\end{table}

\begin{table}
\caption[t8]{The metal system at $z=2.2058$.}
{\label{t6}}
\epsfysize=28cm
\centerline{\epsffile{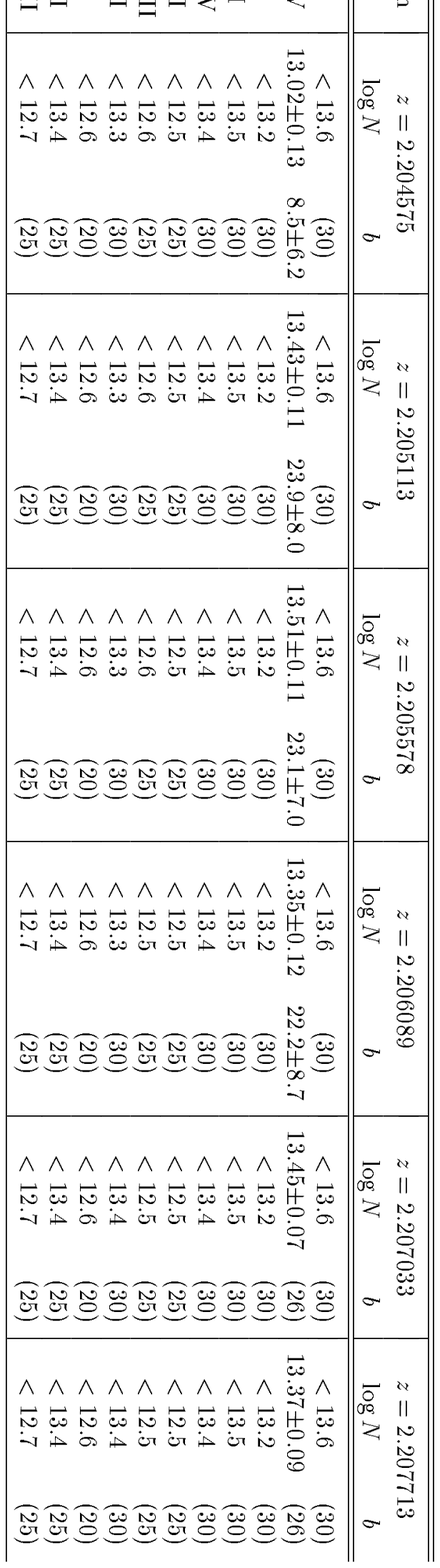}}
\end{table} 

\end{document}


\begin{figure}
\caption[1]{\label{f1} The QSO J2233--606 flux calibrated spectrum
smoothed over 2 \AA. Overplotted is the spectrum of the noise per
pixel. The emission lines correspond to the SiIV doublet ($\sim4400$
\AA), CIV doublet ($\sim5000$ \AA) and AlII+CIII] lines ($\sim6200$
\AA). Ticks indicate absorption lines identified in the
spectrum. The features around 6850 \AA~and 7600 \AA~are due to
atmospheric absorptions of water vapor.}  
\epsfxsize=18cm
\epsfysize=10cm 
\centerline{\epsffile{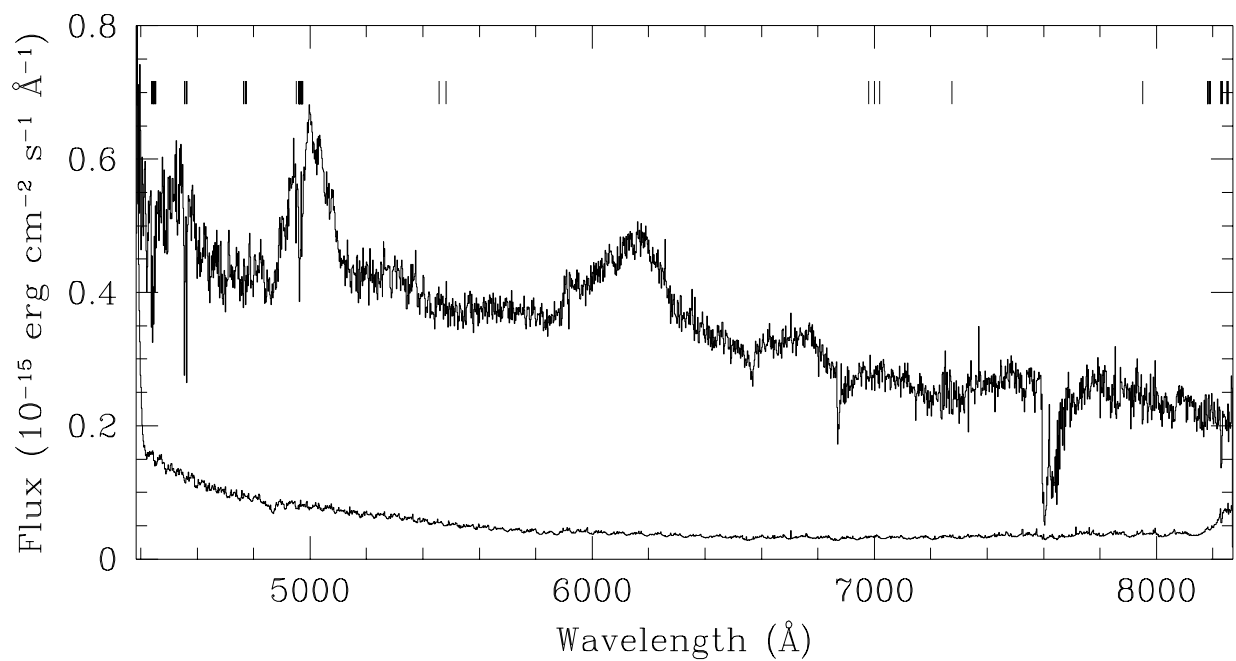}}
\end{figure}

\begin{figure}
\caption[2]{\label{fm0} The MgII absorption of the metal system at
$z=1.5034$. The thin line is the result of the fit. Also shown is the
noise per pixel.}
\epsfxsize=9cm
\epsfysize=12cm
\centerline{\epsffile{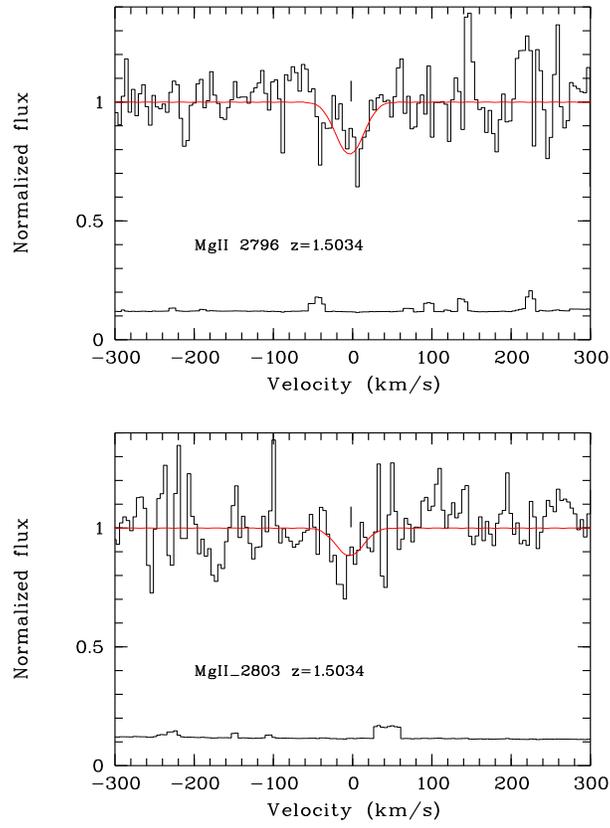}}
\end{figure}

\begin{figure}
\caption[3]{\label{fm0b} As in Fig.~\ref{fm0}, but for the CIV
absorption of the metal system at $z=1.7865$.}
\epsfxsize=9cm
\epsfysize=12cm
\centerline{\epsffile{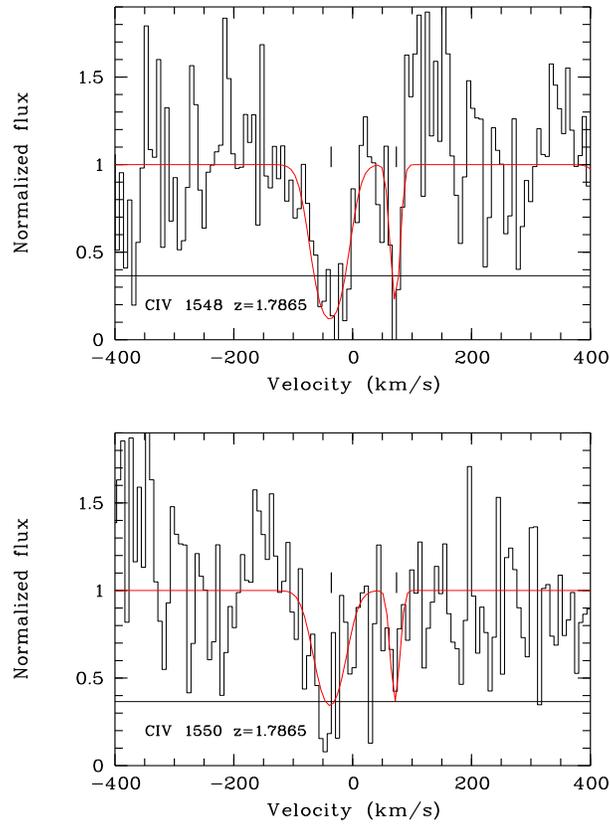}}
\end{figure}

\begin{figure}
\caption[4]{\label{fm1}  As in Fig.~\ref{fm0}, but for the CIV and the
MgI absorption lines of the metal system at $z=1.869$.}
\epsfxsize=9cm
\epsfysize=12cm
\centerline{\epsffile{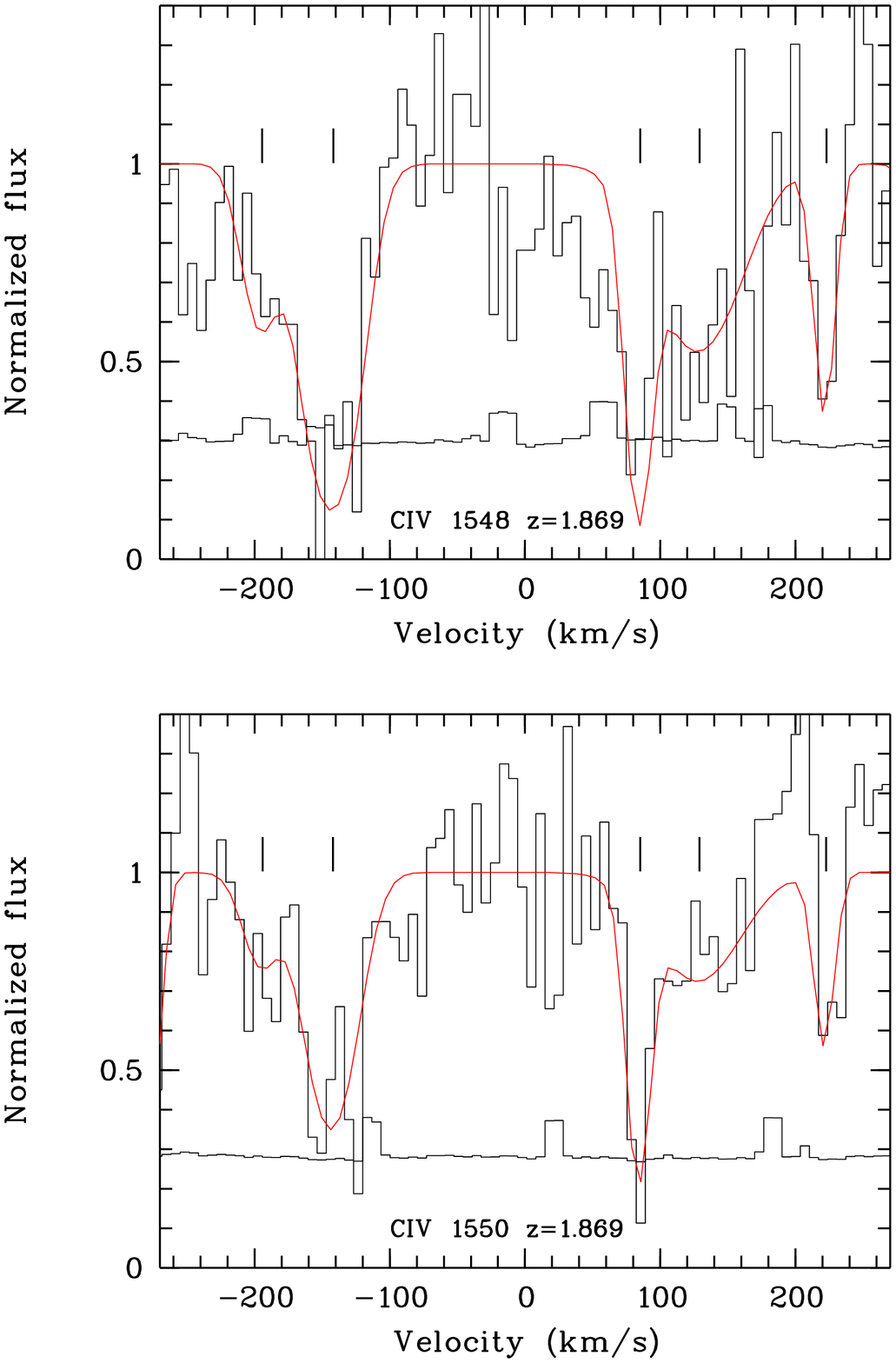}}
\epsfxsize=9cm
\epsfysize=7cm
\centerline{\epsffile{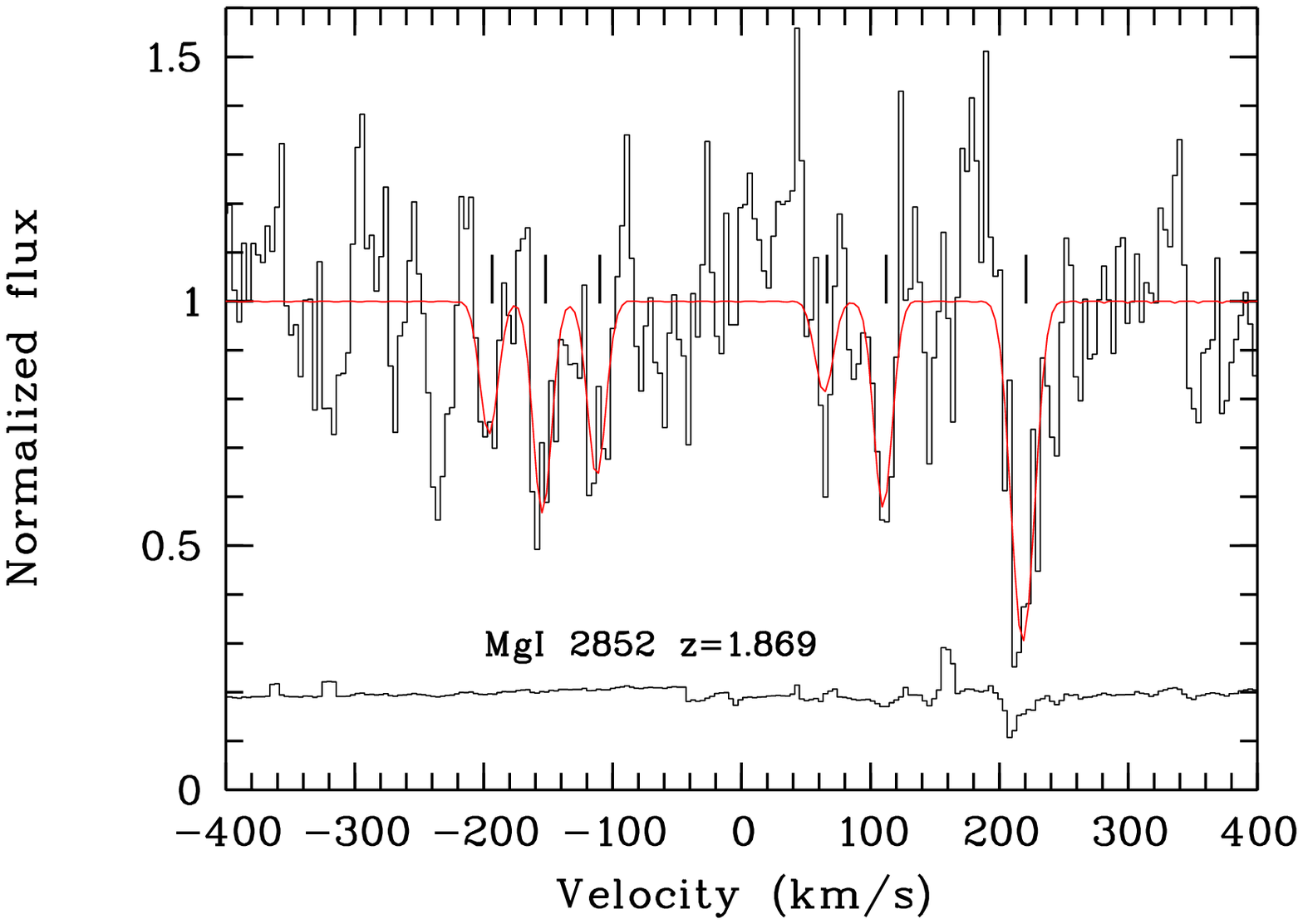}}
\end{figure}

\begin{figure}
\caption[5]{\label{fm2} As in Fig.~\ref{fm0}, but for the CIV and MgII
doublets of the metal system at $z=1.943$.}
\epsfxsize=9cm
\epsfysize=12cm
\centerline{\epsffile{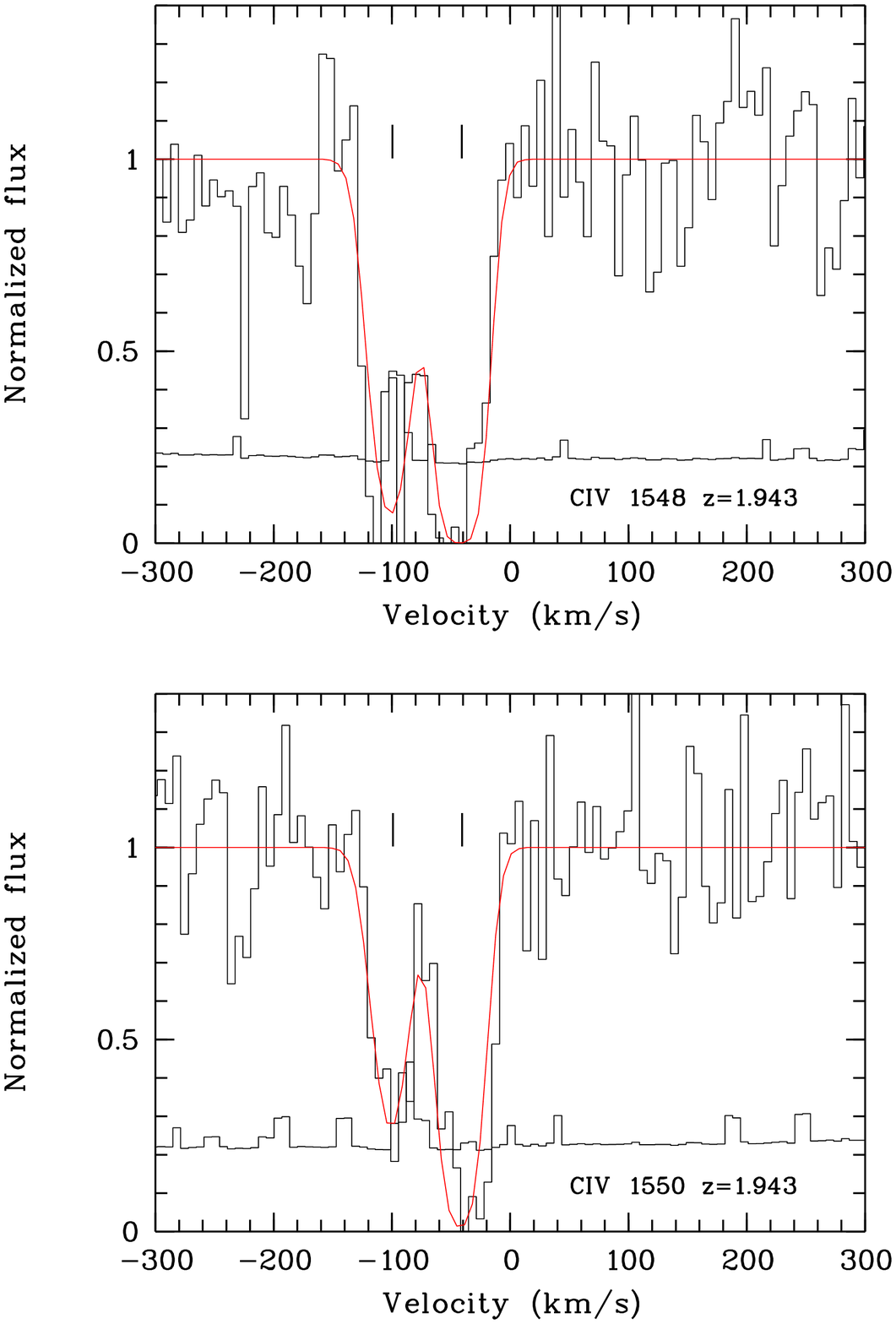}}
\epsfxsize=9cm
\epsfysize=12cm
\centerline{\epsffile{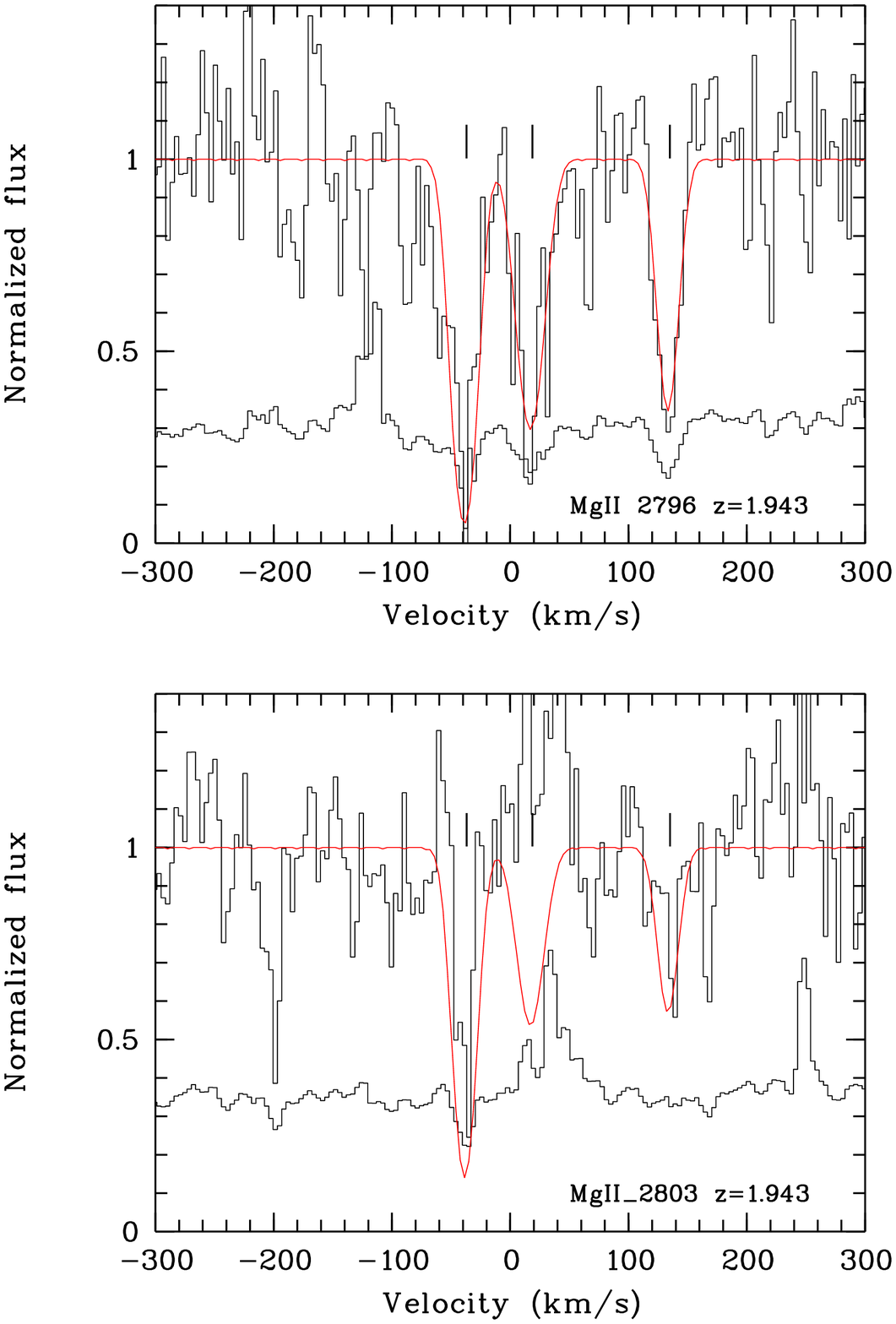}}
\end{figure}

\begin{figure}
\caption[6]{\label{fm3} As in Fig.~\ref{fm0}, but for 
the CIV doublet of the metal system at $z=2.077$.}
\epsfxsize=9cm
\epsfysize=12cm
\centerline{\epsffile{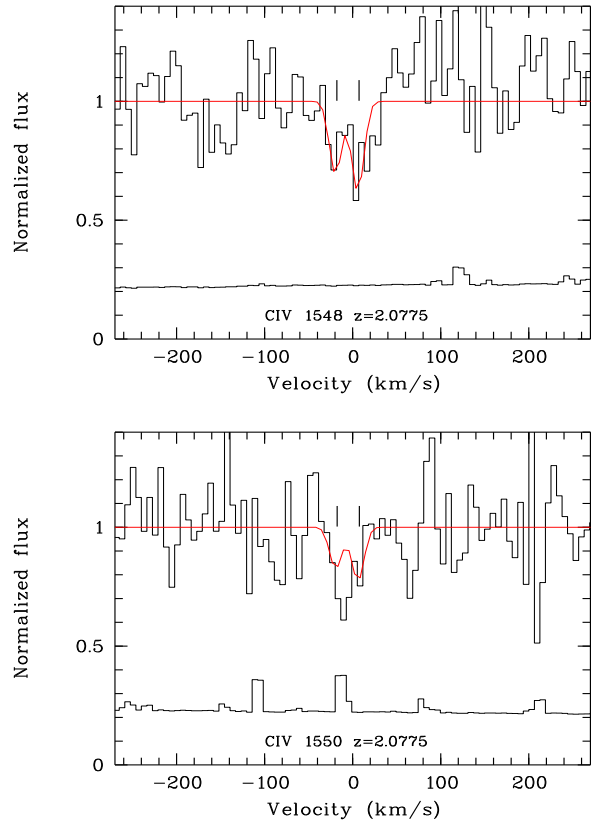}}
\end{figure}

\begin{figure}
\caption[7]{\label{f3} The QSO CIV emission peak. Also shown is the
spectrum of the noise per pixel. }
\epsfxsize=13cm
\epsfysize=7cm
\centerline{\epsffile{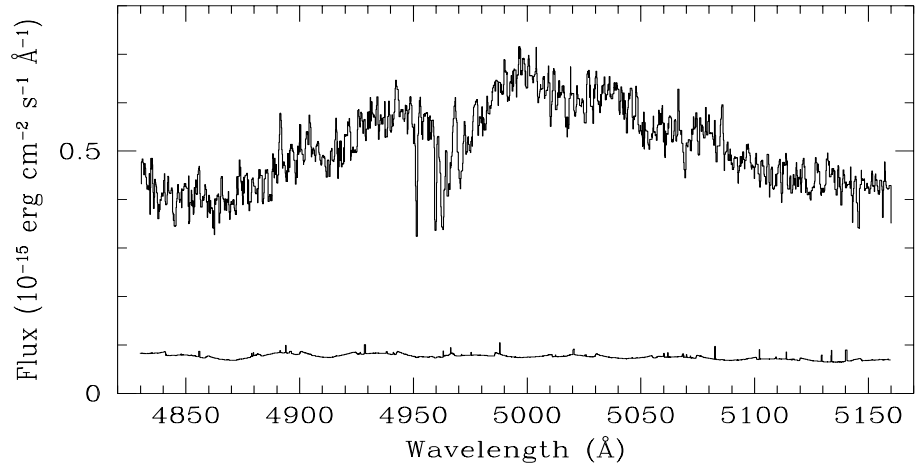}}
\end{figure}

\begin{figure}
\caption[8]{\label{f2} Column density ratios in associated
systems collected from the literature (Table 8). 
The straight lines divide the area of non--detection in case of
observational limits for CIV, SiIV and NV of $\log N_{CIV} = 12.5$,
$\log N_{NV} = 12.1$ and $\log N_{SiIV} = 12.1$ respectively.}
\epsfxsize=15cm
\epsfysize=10cm
\centerline{\epsffile{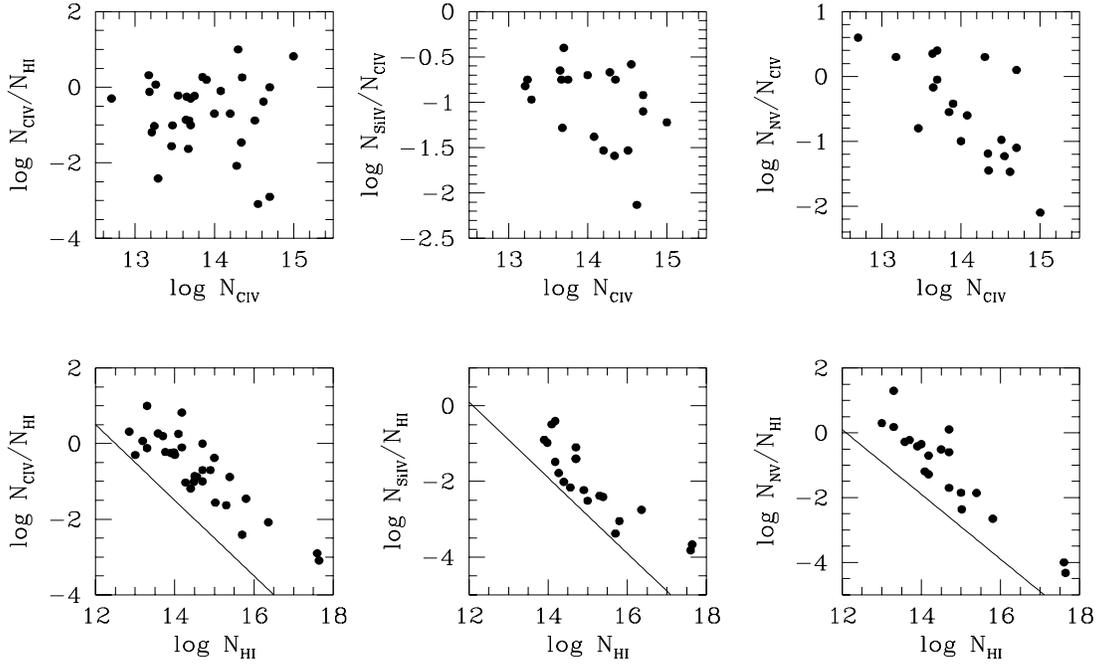}}
\end{figure}

\begin{figure}
\caption[9]{\label{fm4} As in Fig.~\ref{fm0}, but for the CIV complex 
of the metal system at $z=2.1982$.}
\epsfxsize=9cm
\epsfysize=12cm
\centerline{\epsffile{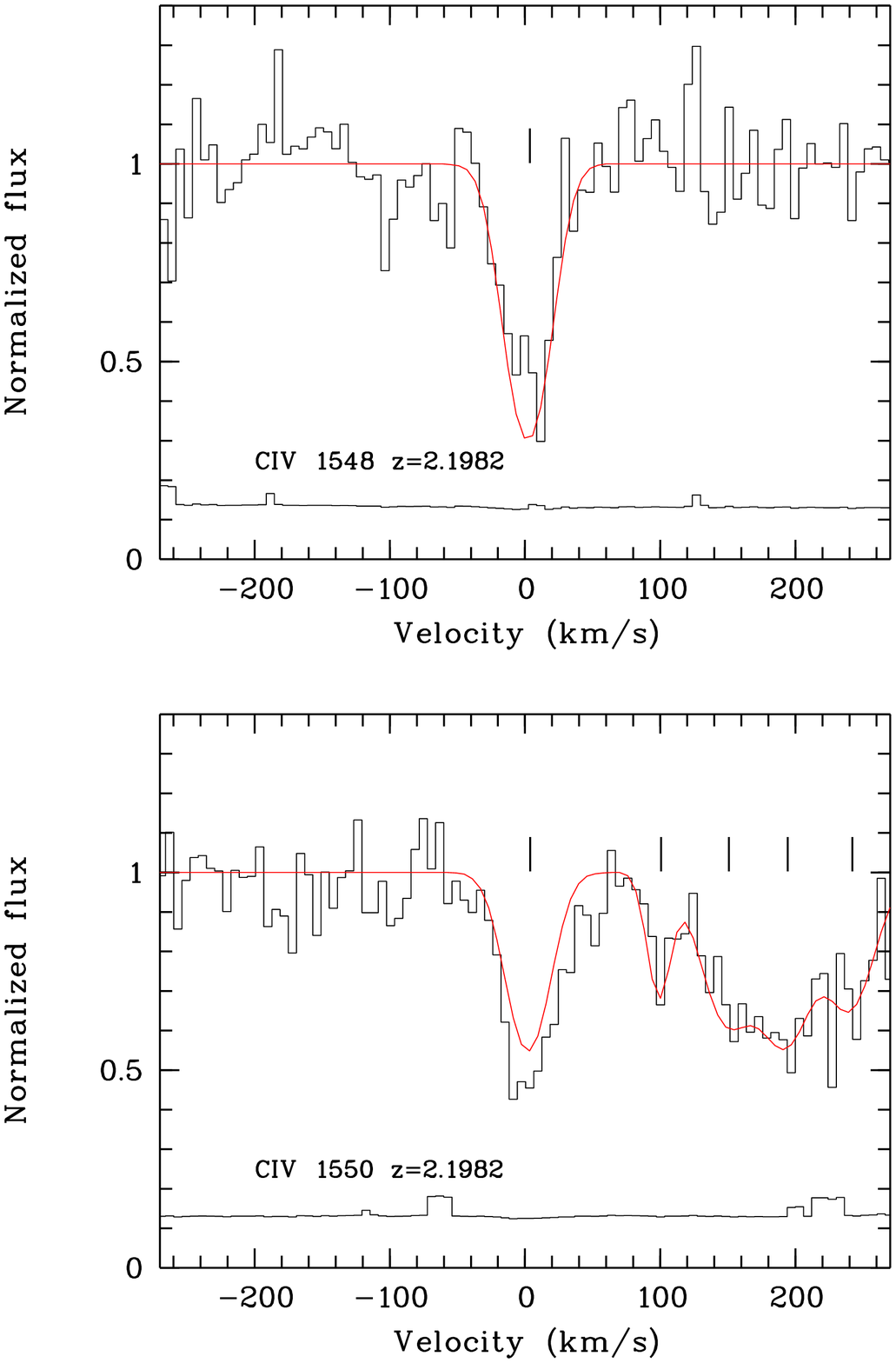}}
\end{figure}

\begin{figure}
\caption[10]{\label{fm5} As in Fig.~\ref{fm0}, but for the CIV complex
of the metal system at $z=2.206$.}
\epsfxsize=9cm
\epsfysize=12cm
\centerline{\epsffile{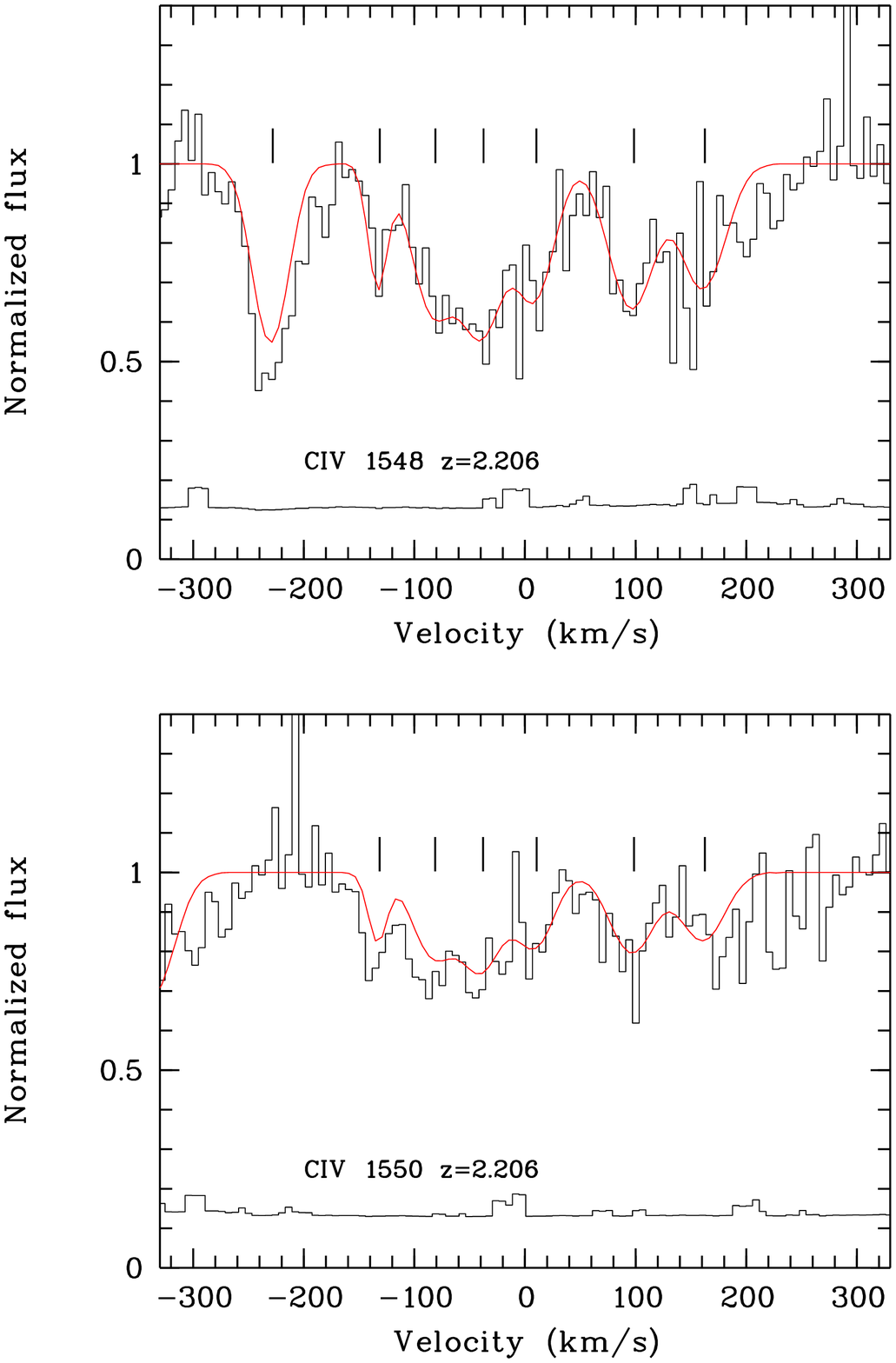}}
\end{figure}

\begin{figure}
\caption[11]{\label{f0} Physical and comoving distances as a function
of redshift  corresponding to  angular separations of 5.1 arcmin
(left vertical axis) and 2.7 arcmin (right vertical axis). The first
is the distance of J2233--606 from the HDFS center, the
second is the angular size of the HDFS. The shaded area indicates the
redshift range of the identified CIV systems in the QSO spectrum.}
\epsfxsize=13cm
\epsfysize=10cm
\centerline{\epsffile{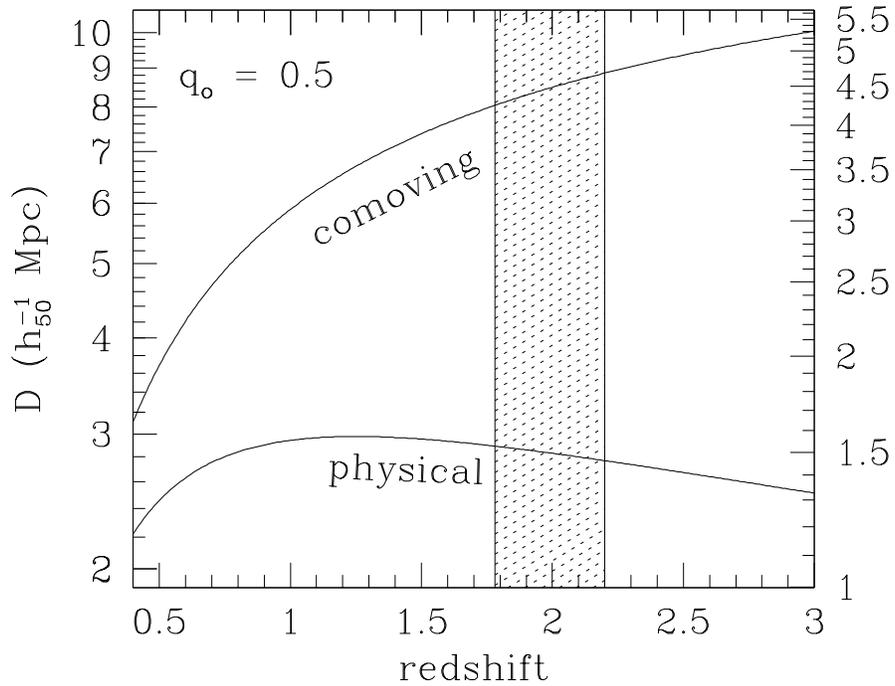}}
\end{figure}